\documentclass[10pt,a4paper,oneside]{report}
\usepackage[T1]{fontenc}

\usepackage{palatino,mathpazo}
\usepackage{amsmath,amsfonts,amssymb}
\usepackage{microtype}
\usepackage{graphicx}
\usepackage[table]{xcolor}
\usepackage{hyperref}
\hypersetup{
			colorlinks = true  , 
			urlcolor   = myBlue, 
			linkcolor  = myRed , 
			citecolor  = myRed   
			}
			
\usepackage{wrapfig}
\usepackage[small,sf,centerlast]{caption}

\usepackage{booktabs}
\usepackage{geometry}
\usepackage{natbib}

\usepackage[scaled=0.88]{beramono}    
\usepackage[scaled=0.90]{biolinum}
\def\MESA{\texttt{MESA}}
\def\code#1{\texttt{#1}}
\def\teff{T_{\rm eff}}

\def\rsol{R_\odot}
\def\msol{M_\odot}
\def\lsol{L_\odot}

\def\llsol{L/L_\odot}

\def\Mbol{\mathrm{M}_{\mathrm{bol}}}

\def\Pg{P_{\mathrm{g}}}
\def\Prad{P_{\mathrm{rad}}}

\def\H1{^1\mathrm{H}}
\def\He3{^3\mathrm{He}}
\def\He4{^4\mathrm{He}}
\def\C12{^{12}\mathrm{C}}
\def\N14{^{14}\mathrm{N}}
\def\O16{^{16}\mathrm{O}}

\def\diff{{\mathrm d}}

\def\unity{ \hbox{1\kern-.23em l} }
\def\zero{ \hbox{0\kern-.23em |} }
\def\field{ \hbox{I\kern-.23em K} }
\newcommand{\afgtitle}[1]{\textsf{\huge #1}}
  
\newcommand{\afgauthor}[1]{{\large{\itshape #1}}}
\newcommand{\afgsection}[1]{{\scshape #1}} 
\newcommand{\afgsubsection}[1]{{\scshape #1}} 

\definecolor{Maroon}{RGB}{128,  0,  0}
\definecolor{myBlue}{RGB}{  8, 85,146}
\definecolor{myRed}{RGB} {138, 10, 11}

\pdfoutput=1
\begin{document}

\centerline{\afgtitle{\`{A} Propos Strange Pulsations of Blue Massive Stars}} 

\centerline{\textcolor{myRed}{\noindent\rule{0.94\linewidth}{0.4pt}}}

\bigskip
\centerline{\afgauthor{Alfred Gautschy}}
\centerline{\textit{CBmA 4410 Liestal, Switzerland}} 

\bigskip

\noindent\makebox[\textwidth][c]{
	\begin{minipage}{0.8\linewidth}
		{\small \noindent
			The properties of radial nonlinear pulsations
			of massive blue stars are computed with
			the \MESA\,software instrument in its
			dynamical mode. Pulsational instabilities
			could be computationally detected and followed
			if the evolutionary timestep was reduced to a
			fraction of the unfolding pulsation period.
			Stellar variability was recovered in regions
			on the HR plane that have been studied before
			and that are known to host LBVs and
			relatives. Mode properties are analyzed on the
			full stellar-evolution models, which are not
			in thermal equilibrium.  Despite persistent
			numerical shortcomings, it appears possible to
			compute strange-mode~--~like pulsations of
			massive blue stars with \MESA.  }
\end{minipage}}

\bigskip\bigskip

\centerline{\afgsection{1. INTRODUCTION}}

The \MESA\,stellar evolution code has proved to be able to capture
sufficiently strong pulsational instabilities in stars (for example
during the early post~--~AGB evolution
\citep{Gautschy2023a} and R CrB~--~like He stars
\citep[][and references therein]{Gautschy2023b} and following some
of them into limit cycles.  If these previously encountered strong and
very rapidly growing pulsational instabilities are indeed connected
with strange modes, then comparable dynamics can be expected to unfold
when the \MESA\,code is used to evolve massive stars on the HR plane
through the S-bend of central hydrogen depletion into core
helium-burning phase when their $L_\ast/M_\ast$ ratios approach and
exceed $10^4$. In nature, this domain is populated with LBV stars
whose cyclic variability is frequently claimed to be associated with
strange modes.  This note presents first positive results of tracking
such pulsations with \MESA\,and it points out the shortcomings that
still prevail in these {\it proof-of-concept }computations.

\bigskip
\centerline{\afgsection{2. EVOLUTION COMPUTATIONS}}

\begin{wrapfigure}{l}{0.40\textwidth}
	\vspace{-20pt}
	\begin{center}{
			\includegraphics[width=0.4\textwidth]{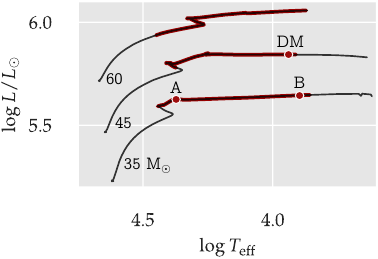}
	}\end{center}
	\vspace{-13pt}
	\caption{Evolutionary tracks of massive stars with superimposed
		instability regions in red. Epochs of steadily pulsating
		model stars, which are analyzed in more detail
		are highlighted and labeled.
	}\label{fig:MassiveHRD}
	\vspace{-10pt}	
\end{wrapfigure}

Stellar models referred to henceforth were computed with
the \MESA\,software instrument in the version close to what is
described in \citet{Paxton2019}.
Methodical and computational aspects of
\texttt{inlist} choices are deferred to Appendix A.
In the attempt to find LBV~--~like pulsating stars with \MESA, massive
star-models (with $60,\,45,$~and $35\,\msol$) were evolved from their
homogeneous ZAMS phase, starting with homogeneous stars with
$Y_{\mathrm{initial}} = 0.26, Z_{\mathrm{initial}} = 0.02$, following
them through central hydrogen burning, and into their early core
helium-burning (Fig.~\ref{fig:MassiveHRD}).  
The evolutionary tracks were computed far enough to
see them enter and cover the region on the HR plane that is aimed at
frequently in LBV studies
\citep[e.g.][and references in the respective articles]
      {dorfiafg00, Saio2009, Sonoi2014, Saio2015, Glatzel2024}. 
In the following, pertinent results obtained along a 
$45\,\msol$ and a $35\,\msol$ track are presented and
discussed.  Suitably tuning the timesteps in \MESA\,(the numerical
setup is roughly outlined in Appendix A) uncovered regions of
oscillatory instabilities for all three masses considered here. The
extent of the instability regions on the HR plane agree decently well
with what other researchers have reported hitherto.  It is important
to emphasize that the details, mainly the magnitude of the growth
rates of the pulsations and hence to some extent the precise blue and
red instability boundaries and the amplitudes of the light and radius
variability depend on the numerical treatment.  Because this is a
proof-of-concept report, we do not attempt to compare in detail the
results with what is observed; it is too early to try to match
numerical results with properties of observed stars.  On the
other hand, the current results are not numerical artifacts and hence
we proof that \MESA\,has the potential to pave the way to better
understand these massive pulsators with the help of a 
stellar-evolution code alone.  Even at this early stage we can glimpse at some
rewarding physics to which we have no access to in the linear regime.

The evolution of model stars with $35, 45$, and $60\,\msol$ with $Z =
0.02$ was, in a first turn, computed (dynamically) with timesteps on
nuclear/thermal timescales; they varied between $10^4$~years during
early main-sequence evolution and some $10$~years during early
core He-burning.  So called \texttt{photo}~models were stored
frequently along the evolutionary track. In a second step, evolution
computations were restarted from \texttt{photo}~snapshots requesting
maximum timesteps of the order of one day or shorter. In contrast to
the computations reported on earlier \citep{Gautschy2023b,
Gautschy2023a}, no regridding was imposed on the initial models of the
small-timestep re-starts.  Instabilities developed at suitable
evolutionary epochs anyway.  Regridding tended to actually disturb the
apparently already delicate hydrostatic equilibria of the stars,
causing the code to frequently stall.  A collection of other tweaks
that proved helpful to compute red-supergiant pulsations
with \MESA\,were put forth e.g. in \citet{Bronner2025}.
The heavy red lines in Fig.~\ref{fig:MassiveHRD} that are superimposed
on the evolutionary tracks highlight the instability regions of the
three model series. The locations of the blue and
the red edges depended somewhat on the magnitude of the maximum timestep
allowed in the computations.  

The $60\,\msol$ model star showed the
onset of oscillatory instabilities while still burning hydrogen in the
core (at about $\log \teff = 4.4$).  In all considered cases, the
instabilities were so strong that \MESA\,stalled or failed to converge
after a few initial cycles because of the onset of violent dynamics in
the star's outermost layers. Compared with the two smaller-mass model
sequences, the total kinetic energy in the pulsations of the
$60\,\msol$ models was about a hundred times higher.  Therefore, for
the present goals, the behavior of the $60\,\msol$ model star is not
further commented on here because it does not contribute to better
understand mode physics at the current stage of the computing affairs.
For the lower two masses, on the other hand, regions of saturation of
the pulsation growth were encountered in the neighborhood of the
instability boundaries. This cyclic variability is taken advantage of
in the following as it allows to address aspects of pulsation-mode
physics.

\bigskip\medskip
\centerline{\afgsection{3. PULSATION PHENOMENOLOGY}}

Three cases of well developed nonlinear pulsations are picked out in
the following.  The corresponding evolutionary epochs are marked along
the respective tracks on the HR plane in Fig.~\ref{fig:MassiveHRD} as
cases "DM", "A", and "B".  The pulsation modes that were encountered
in this exercise are likely the coolward and hence longer-period
extension of the strange modes discussed for example
in \citet{dorfiafg00}. The cycle-lengths of the bluemost pulsations of
the \MESA\,computations are somewhat longer than the periods reported
in Dorfi \& Gautschy. This is not surprising because overall the
\MESA\,evolutionary tracks are cooler than those in the old computations.

\bigskip\medskip
\centerline{\afgsubsection{3.1. $45\,\msol$ Star: Case DM}}

The highlighted epoch at $\log \teff = 3.94,\,\log \llsol = 5.84$,
referred to as case DM, lies close to the red edge of the instability
region of the $45\,\msol$~model sequence. The temporally
high-resolution run limited the timestep
via \code{max$\_$years$\_$for$\_$timestep = 1.0d-3}~years ($\approx
0.37$~days).
\begin{wrapfigure}{r}{0.65\textwidth}
	\vspace{-0pt}
	\begin{center}{
			\includegraphics[width=0.65\textwidth]{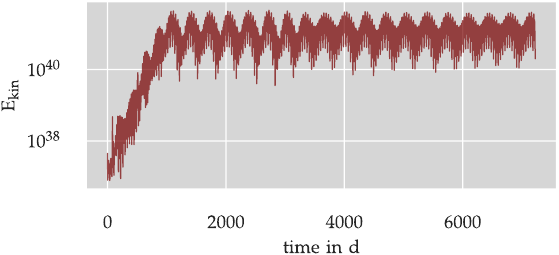}	
	}\end{center}
	\vspace{-15pt}
	\caption{The very steep growth of the oscillatory variation of
                 the total kinetic 
		 energy (in erg/s) of the model star saturates after
		 about $1000$~days the simulation~--~that is,
                 after about $30$~cycles. 
	}\label{fig:M450Z02_Saturation}
	\vspace{-30pt}	
\end{wrapfigure}

The emerging growth rate of the pulsation kinetic energy shown in
Fig.~\ref{fig:M450Z02_Saturation} is a typical example of the speed of
pulsation growth that was encountered in the models in the instability
regions traced out in Fig.~\ref{fig:MassiveHRD}.

The amplitude modulation of the temporal evolution of the kinetic
energy of model DM indicates that more than one pulsation mode is
involved.  A Lomb-Scargle analysis revealed two periods being
present: The dominant (with respect to amplitude) mode has a period of
$38.9$~days, the second one has a period of $43.8$~days.

\begin{wrapfigure}{l}{0.40\textwidth}
	\vspace{-10pt}
	\begin{center}{
			\includegraphics[width=0.40\textwidth]
			{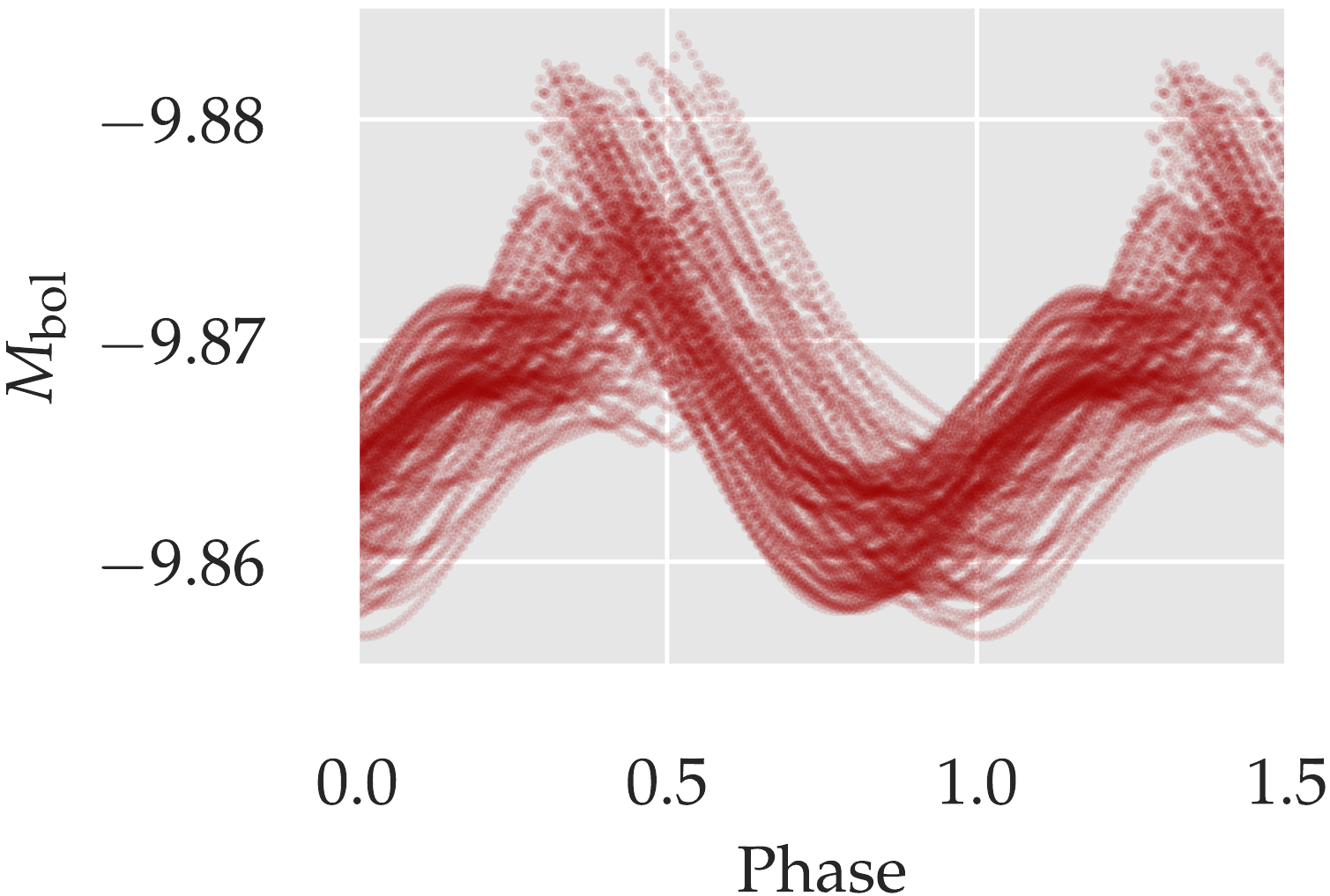} 
	}\end{center}
	\vspace{-15pt}
	\caption{The bolometric light variability of 
		the double-mode pulsating model star folded with 
		the dominant periodicity of $38.9$~days.
	}\label{fig:M450Z02_bolometricLight}
	\vspace{-30pt}	
\end{wrapfigure}
Within $1000$~days, the kinetic energy of the developing pulsation
growed to about $10^{41}$~erg (Fig.~\ref{fig:M450Z02_Saturation}) when
it saturated.  The corresponding e-folding time is about
110~days. Hence, the amplitude grows by a factor of e within about
three to four cycles; such numbers are reminiscent of strange modes.

The slopes of the instability growth-rates depended on the choice of 
\texttt{max\_years\_for\_timestep}. The larger the maximum timestep was the 
slower the growth.  This behavior hints at numerical dissipation to be
responsible for the problem.  Nonetheless, in all cases where
instabilities could be tracked, the e-folding times remained always of
the order of a few pulsation cycles.

The phase plot of Fig.~\ref{fig:M450Z02_bolometricLight} consists of
the bolometric lightcurve over $122$~cycles that was folded by the
dominant mode period of $38.9$~days. Whether the double-mode behavior
is persistent in the long run is unclear. The slightly shrinking
amplitude of the modulation in Fig.~\ref{fig:M450Z02_bolometricLight}
might indicate a transient phenomenon. In any case, this aspect is
beyond the scope (and CPU accessibility) of this report and the
encountered phenomenology is beautiful enough to justify its showing
off. Actually, the double-mode behavior permits a glimpse at a small
range of the star's mode spectrum: The deduced period ratio of $0.89$
of the two involved pulsation modes is unusually large when compared
with classical pulsators.  The ratio of adjacent periods of
(adiabatic) pressure modes of classical pulsators is closer to 0.75.
On the other hand, stars hosting strange modes are prone to mode
interaction \citep[e.g.][]{Saio2011}. Therefore, the period ratio is
not that surprising if strange modes are involved in model DM.

The computed amplitude of the bolometric light variation is likely
affected by numerical dissipation in the \MESA\,coding too. 
The currently reported amplitudes are therefore lower than what is to
be expected in nature. Nonetheless, the $\Delta \Mbol = 0.02$
from Fig.~\ref{fig:M450Z02_bolometricLight} is not in conflict with 
observed photometric small-scale variability of massive blue stars.

\bigskip\medskip
\centerline{\afgsubsection{3.2. $35\,\msol$ Star: Case A}}

\begin{wrapfigure}{r}{.45\textwidth}
	\vspace{-23pt}
	\begin{center}{\includegraphics[width=0.45\textwidth]{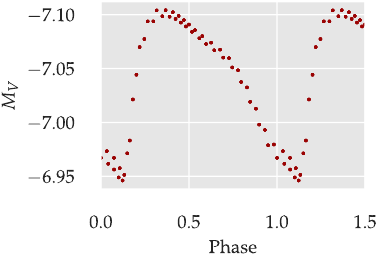}
	              }\end{center}
	\vspace{-15pt}
	\caption{The photometric V-band lightcurve of case-A model
                 star folded with the period of 1.79~days. 
	}\label{fig:M350Z02_CaseA_VLight}
	\vspace{-20pt}	
\end{wrapfigure}

Case A in Fig.~\ref{fig:MassiveHRD} at $\log \teff =
4.38,\,\log \llsol = 5.65$ models a $35\,\msol$ star during its early
post~--~main-sequence evolution phase when all nuclear energy is
generated in a thick hydrogen-burning shell. The case A model is close
to the blue edge of instability region and still rather compact, not
far from thermal equilibrium; it pulsates with a period of
$1.79$~days.  The pulsation growth is comparable to what is displayed
in Fig.~\ref{fig:M450Z02_Saturation} or
Fig.~\ref{fig:M350Z02_Saturation}.  The mean kinetic energy reached
$\approx 4\times 10^{41}$~erg once the pulsation saturated.

The Johnson-Cousins V-band ($M_V$) lightcurve has an amplitude of
$0.15$~magnitudes (Fig.~\ref{fig:M350Z02_CaseA_VLight}). Each point
plotted in the phase plot is actually a superposition of $91$ distinct
epochs from the simulation window that were folded by the star's
pulsation period.  Accidentally, the time steps of
the \MESA\,computations and the model star's pulsation period were
commensurable. 

\begin{wrapfigure}{l}{.45\textwidth}
	\vspace{-25pt}
	\begin{center}{\includegraphics[width=0.45\textwidth]{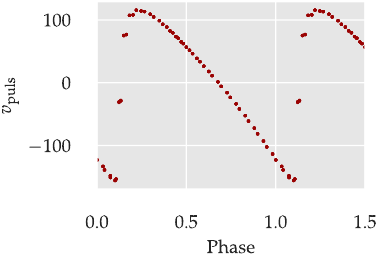}
	              }\end{center}
	\vspace{-15pt}
	\caption{Case-A pulsation velocity curve folded
		with the period of 1.79~days. 
	}\label{fig:M350Z02_CaseA_vpuls}
	\vspace{-20pt}	
\end{wrapfigure}
The small size of the points in the phase diagram
indicate the good stability of the pulsation period over the timespan
considered. The dots marking the lightcurve hint also at two distinct
lightcurve shapes with equal period but differing slightly in their
amplitudes. Lightcurves spanning several cycles reveal that the
amplitude of the photometric light variability changes by $6 \times
10^{-3}$~mag between two successive cycles.

In contrast to the photometric lightcurve, the pulsation-velocity
variation was almost devoid of the large/small amplitude
alternation. Even though the photometric amplitude is small, that of
the associated pulsation velocity is huge: It ranges from $-157$ to
$+115$~km/s.  The star's radius varies accordingly between $34$ and
$42.5\,\rsol$; this means that the relative (to minimum radius)
variation amounts to $25\,\%$.  Despite the expressed nonlinearity of
the pulsation, the initial "equilibrium" radius of the evolutionary
model of $37.7\,\rsol$ is only slightly smaller than the arithmetic
mean radius of the pulsating star.
According to Figs.~\ref{fig:M350Z02_CaseA_VLight} and 
\ref{fig:M350Z02_CaseA_vpuls} there is~--~in contrast to classical 
pulsators~--~essentially no phase lag, for example between minimum
V-light and minimum radius. This aspect will be expanded upon in the
next section for further observables.

\bigskip\medskip
\centerline{\afgsubsection{3.3. $35\,\msol$ Star: Case B}}

Close to the red edge of the instability domain of the $35\,\msol$
model sequence, another model star could be tracked down that developed 
a limit-cycle type pulsation.  The star standing for case B in
Fig.~\ref{fig:MassiveHRD}, at $\log \teff = 3.91,\,\log \llsol =
5.64$, is already in early core He-burning and considerably out of
thermal equilibrium. In the temporally high-resolution run discussed
in the following, the timestep was limited to
\code{max$\_$years$\_$for$\_$timestep = 3.0d-4}~years ($\approx 0.1$ days).

\begin{wrapfigure}{r}{0.65\textwidth}
	\vspace{-20pt}
	\begin{center}{\includegraphics[width=0.65\textwidth]{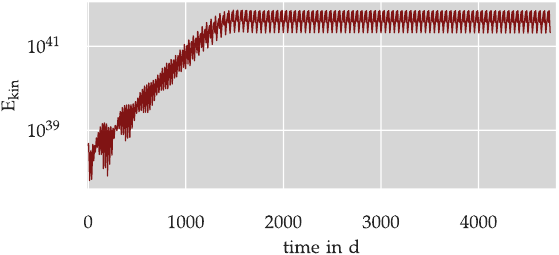}
	              }\end{center}
	\vspace{-15pt}
	\caption{Exponential growth and saturation of case B's total 
		kinetic energy (in erg).
	}\label{fig:M350Z02_Saturation}
	\vspace{-10pt}	
\end{wrapfigure}
\begin{wrapfigure}{l}{0.35\textwidth}
    \vspace{-20pt}
	\begin{center}{\includegraphics[width=0.35\textwidth]{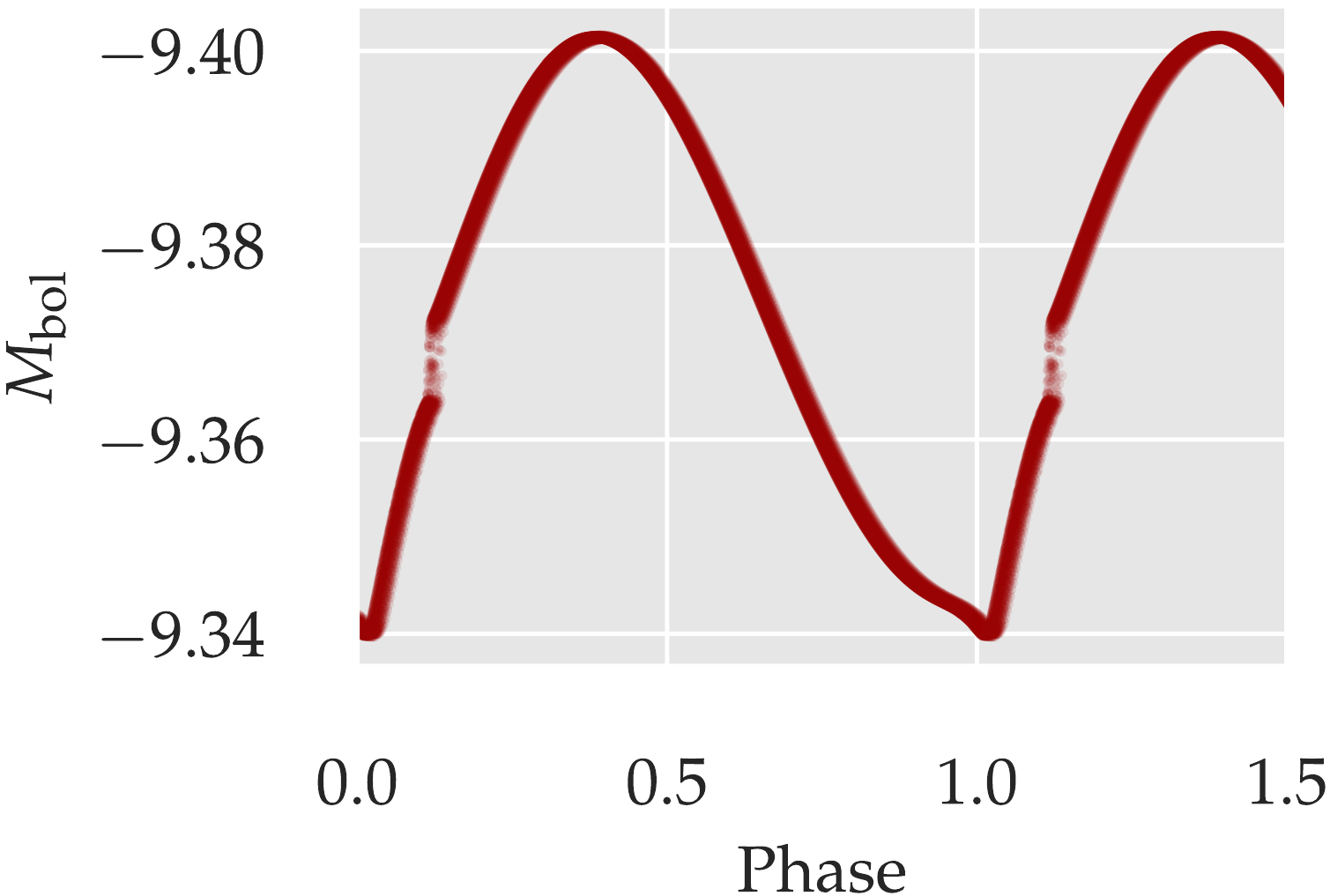}
                      }\end{center}
	\vspace{-15pt}
	\caption{The bolometric light variability as 
			 obtained from temporal data folded with the pulsation 
			 period of $45.64$~days. 
    	    }\label{fig:M350Z02_Bolometriclight}
	 \vspace{-20pt}	
\end{wrapfigure}
\begin{wrapfigure}{r}{0.4\textwidth}
	\vspace{-30pt}
	\begin{center}{\includegraphics[width=0.4\textwidth]{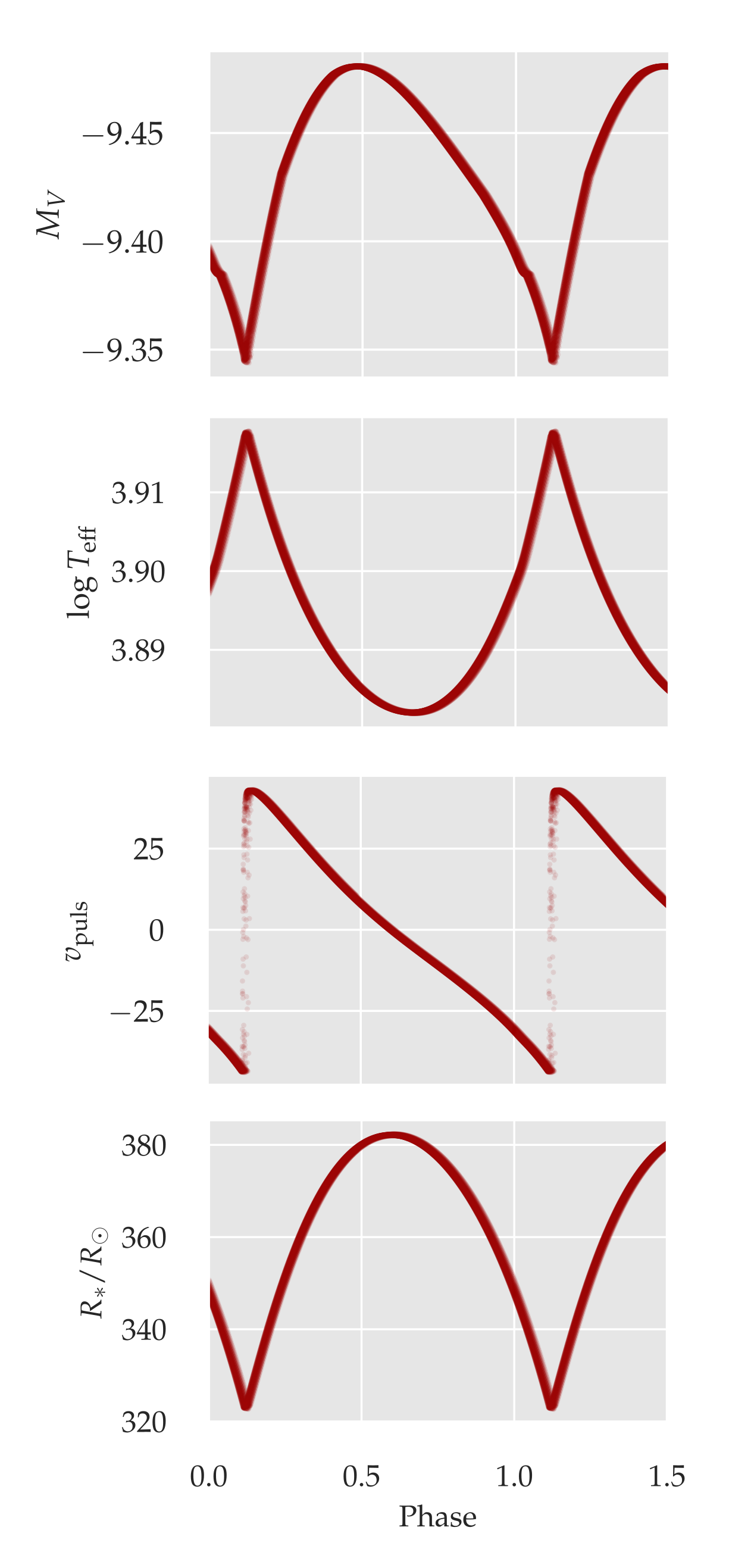}
	              }\end{center}
	\vspace{-15pt}
	\caption{Phase plots of some observables of case B model 
		illustrate the lack of phase shifts between them.  
	}\label{fig:M350Z02_VariousPhasePlots}
	\vspace{-50pt}	
\end{wrapfigure}

After about 1500~days of model time, the pulsation
that developed out of random numerical noise, saturated above 
$10^{41}$ erg (Fig.~\ref{fig:M350Z02_Saturation}), 
at a somewhat higher kinetic energy level than the $45\,\msol$ case .
The mono-mode behavior that eventually developed can be regarded as a good
approximation to a cyclic pulsation.  At around $5000$~days into the
simulation, the last $36$~cycles in the computational run were selected 
to go into the phase plots shown in Figs.~\ref{fig:M350Z02_Bolometriclight} 
and \ref{fig:M350Z02_VariousPhasePlots}. The quality of the stacking
of the involved cycles proves that the derived pulsation period of
$45.64$~days was already quite robust.

The variability of the bolometric brightness makes for a slightly 
asymmetric lightcurve (Fig.~\ref{fig:M350Z02_Bolometriclight}). The
ascending branch is shorter than the descending one, which
develops a weak bump just before minimum light. The brightness-jump on
the ascending branch is physically understandable but possibly only a
numerical artifact not observable in nature: The jump is inflicted by 
a brief interference with the photosphere of the very sharp opacity bump 
caused by partial hydrogen ionization. 

The top panel of Fig.~\ref{fig:M350Z02_VariousPhasePlots} shows the 
photometric light variability in the V-band. 
The photometric lightcurve has a larger amplitude than the bolometric one, 
and the discontinuity that the latter exhibited on the rising branch disappeared. 
In fact, the phase of the jump in bolometric light coincided with 
the unusually sharp minimum in the V-band. 

The dynamics of the outermost stellar layers remains periodic and
well behaved over many cycles. The lowest two panels of
Fig.~\ref{fig:M350Z02_VariousPhasePlots} illustrate this superficial
dynamics: The pulsation velocity at the stellar surface,
$v_{\mathrm{puls}}$, varied between $-44$ and $+43$
km/s. The quick transition from maximum infall- to maximum expansion-speed 
is of course also reflected in the star's radius variation shown in the
bottom panel. The form of both curves hints at an abrupt bounce of the
photospheric layers followed by an ensuing essentially ballistic
movement. The relative stellar-radius variation, $\Delta R /
R_{\mathrm{min}}$ amounts to $19 \%$ as the photosphere oscillates
between $322$ and $382\,\rsol$. 

The most striking properties of the phase-aligned observables shown in 
Fig.~\ref{fig:M350Z02_VariousPhasePlots} are the matching of minimum
light (in $M_V$) at bounce phase (i.e. at the short phase-window in
which the pulsation speed snaps from maximum infall to maximum
expansion) so that minimum light coincides essentially with minimum
radius when the photosphere is hottest. Hence, there is \emph{no
}phase shift between minimum $V$-light, minimum radius, and maximum
effective temperature. This differs significantly from what is
observed in classical pulsators such as Cephe\"ids for which maximum
expansion speed and maximum surface temperature are reached around
maximum light. Phenomenologically, a Cephe\"id's lightcurve looks 
more like a slightly distorted surface-temperature curve. 
In contrast, the case-B pulsation data collected in
Fig.~\ref{fig:M350Z02_VariousPhasePlots} show that the $V$-band
lightcurve is reminiscent of the radius-variation curve.

\medskip

\hfil{4. MODE PHYSICS}\hfil

From Fig.~\ref{fig:M350Z02_VariousPhasePlots} it becomes clear that
the late phases of the simulation of the $35\,\msol$ model star
approximate a limit cycle pretty well. In this section now,
integrations over such late cycles of case B model (see
Fig.~\ref{fig:M350Z02_Saturation}) are used to gain insight on the
physical properties of the excited pulsation mode. The modal behavior
encountered in case B is mostly representative of what was seen in other
cyclically pulsating stars of the model sequences that constitute
Fig.~\ref{fig:MassiveHRD}.

\begin{wrapfigure}{l}{0.48\textwidth}
	\vspace{-20pt}
	\begin{center}{\includegraphics[width=0.48\textwidth]{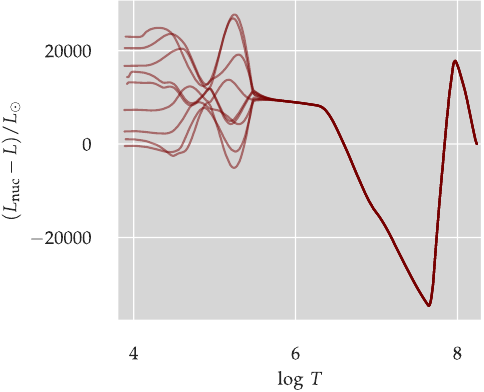}
		      }\end{center}
	\vspace{-15pt}
	\caption{Superposition of a few phases during a 
		pulsation cycle of the difference between the luminosity 
		from nuclear burning and the radiated one of case B model. 
	}\label{fig:M350Z02_LuminosityImbalance}
	\vspace{-20pt}	
\end{wrapfigure}

Selected phases during a pulsation cycle of the difference between
nuclear ($L_{\mathrm{nuc}}$) and radiated luminosity ($L$) profiles
(Fig.~\ref{fig:M350Z02_LuminosityImbalance}) highlight interesting
properties: In the envelope (say layers where $\log T \lesssim 6$),
$L_{\mathrm{nuc}} - L$ varied cyclically with the model star's
pulsation period. This short-term variability is quenched around $\log
T = 5.7$. Down to about $\log T = 5$, its spatial variation looks much
like a standing wave. At even lower temperatures, the variability
eventually picked up a running-wave component.  The pulsations reported
here were encountered on complete evolutionary model stars. In
particular, case B in its early core He-burning stage with a still
dominant H-burning shell at the bottom of the envelope is not in
thermal equilibrium.  The high-temperature part ($\log T \gtrsim 6$)
of Fig.~\ref{fig:M350Z02_LuminosityImbalance} shows a
thermal imbalance contribution that exceeds the one from the
pulsation. Above the He-burning core, the luminosity from the
respective nuclear burning exceeds the radiated one. Across and above
the H-burning shell (to about $\log T = 6.6$), the radiated luminosity
exceeds the nuclear one. Finally, the nuclear inert envelope radiates,
once again, less than what nuclear processes provide.

Stellar energetics relates the rate of change of the 
integrated heat content of stellar matter, $\hat{q}(m)$, to the star's 
thermal imbalance:
$$
\frac{\diff \hat{q}(m)}{\diff t} = L_{\mathrm{nuc}}(m) - L(m) \,.
$$
The above relation means that the case B model star gains heat above 
the He-burning core and in the envelope but loses heat in the mantle between the 
base of the envelope and the H-burning shell. The zeros of 
$L_{\mathrm{nuc}}(m) - L(m)$ in the deep interior mark the nodes of the
radius evolution of the star in thermal imbalance.  
 
To determine the average net rate at which the star would gain heat
if this heat were not being transformed into pulsation energy,  
the mass-integrated heat-gain/-loss rate be further integrated over a
pulsation period $\Pi$:
\begin{equation}
	\oint_\Pi \diff \hat{q}(m) \equiv
	\oint_\Pi \frac{\diff \hat{q}(m)}{\diff t} \diff t 
	\varpropto \oint_\Pi  \diff t
	            \int_0^m P(m')  \frac{\diff V(m')}{\diff t}\, \diff m' 
	\label{eq:integrated_heat}
\end{equation}
with the last integral being the total (from stellar center to $m$)
work done per cycle by the stellar material on its surroundings.
Because $\oint_\Pi \diff u(m)$, with $u$ being the specific internal
energy, fails to zero out~--~mostly in the outermost regions where the
pulsation has considerable traveling-wave character~--~only a
proportionality can be written rather than an equality. Nonetheless, 
the cycle integrated heat-gain/-loss remains a good indicator of the 
work integral to which stellar pulsation theory frequently resorts to.

\begin{wrapfigure}{r}{0.45\textwidth}
	\vspace{-20pt}
	\begin{center}{\includegraphics[width=0.45\textwidth]{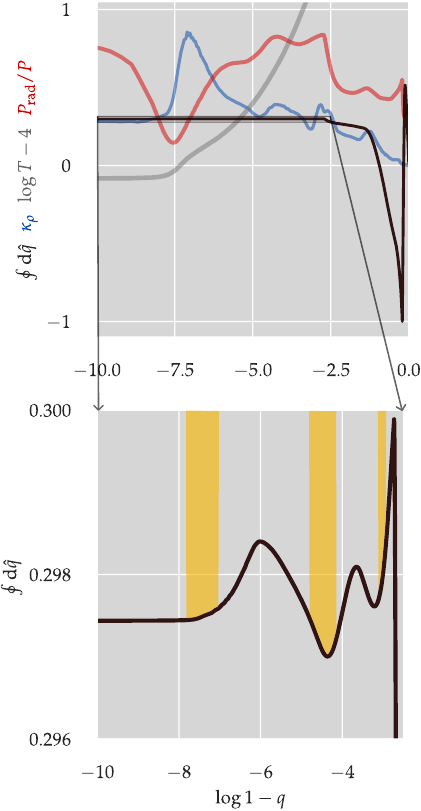}
	              }\end{center}
	\vspace{-15pt}
	\caption{The thermal imbalance profile integrated over a pulsation cycle 
		(black) and physical structure profiles at the epoch of the 
		beginning of the cycle (around minimum radius) of case-B model. 
	}\label{fig:M350Z02_CycleIntegratedHeat_and_StructuralProperties}
	\vspace{-30pt}	
\end{wrapfigure}

The black line in the top panel of
Fig.~\ref{fig:M350Z02_CycleIntegratedHeat_and_StructuralProperties}
shows the left-hand side of eq.~\ref{eq:integrated_heat} evaluated
over a late pulsation cycle of case-B model. For ease of stacking with
other quantities, the $\oint_\Pi \diff \hat{q}$ profile is scaled to its
most pronounced feature; that means, to the magnitude of the minimum
reached on top of the H-burning shell.  The choice of $\log(1 - q)$ as
abscissa (with $q\equiv m/M_\ast$) conveniently stretches the envelope
but compresses the central regions. The quasi-static evolutionary
thermal imbalance contributions that are well expressed on the right
half of Fig.~\ref{fig:M350Z02_LuminosityImbalance} are still
recognizable in
Fig.~\ref{fig:M350Z02_CycleIntegratedHeat_and_StructuralProperties}
but they are huddled together against the figure's right edge.

To facilitate the identification of $\oint_\Pi \diff \hat{q}$ features in
the stellar interior, the top panel of
Fig.~\ref{fig:M350Z02_CycleIntegratedHeat_and_StructuralProperties}
contains additional useful stellar-structure profiles derived from the
starting model of the respective pulsation cycle. The start phase of a
pulsation cycle was chosen to be at the pointed minimum radius.  The
grey curve traces the run of temperature in the star's envelope.  The
red profile shows the variation of the radiation pressure relative to
the total pressure: The inner envelope, between $10^5$ and $10^6$~K,
is clearly dominated by radiation pressure; there it can contribute up
to $80\,\%$ of the total pressure. The full blue line,
$\kappa_{\rho} \equiv \partial\log \kappa
/ \partial \log \rho\vert_T$, is a tracer of the opacity behavior
throughout the star; its bumps correlate with $\diff\,\kappa_T
/ \diff\,r >0$ (highlighted by yellow bars in the bottom panel of
Fig.~\ref{fig:M350Z02_CycleIntegratedHeat_and_StructuralProperties}),
which are associated with driving by the canonical $\kappa$-mechanism.

The quasi-static contribution to $\oint_\Pi \diff \hat{q}$ of the evolution-induced 
thermal imbalance means that the former just accumulates by an integration 
over a pulsation cycle. The situation in the envelope (say $\log T \lesssim 6$, 
i.e. $\log (1-q)\lesssim -2$) is different though. 
At first sight, the considerable $\diff\hat{q}(m) / \diff t$ variability 
that is evident in Fig.~\ref{fig:M350Z02_LuminosityImbalance} 
seems to cancel over a pulsation cycle (top panel of
Fig.~\ref{fig:M350Z02_CycleIntegratedHeat_and_StructuralProperties}). 
Only a zoom-in reveals that pulsation-induced heat gains and losses
persist on a $10^{-3}$ level as compared with the evolutionary ones
(lower panel of Fig.~\ref{fig:M350Z02_CycleIntegratedHeat_and_StructuralProperties}).
Regions of heat gain (driving pulsations) are characterized by 
negative slopes of the black line in the lower panel of
Fig.~\ref{fig:M350Z02_CycleIntegratedHeat_and_StructuralProperties}. The
rightmost, deepest lying spike is not inflicted by the pulsation but it is 
already present in the quasi-hydrostatic evolution computed at large timesteps. 

Potential driving regions by means of the canonical $\kappa$-mechanism
can be identified via $\kappa_{\rho}$ bumps in the top panel of 
Fig.~\ref{fig:M350Z02_CycleIntegratedHeat_and_StructuralProperties} and
even more so via the yellow bars in the bottom panel that mark the 
regions where $\diff\,\kappa_{T}/\diff\,r > 0$. 
From left to right, they are associated with
H/HeI, HeII partial ionization, and the deepest-lying one with the $Z$-bump
in the Rosseland opacity profile. The slopes of the 
$\oint_\Pi \diff \hat{q}$ profile zoomed into in the bottom panel 
show at best partial correlation only with the requirements for 
the action of the classical $\kappa$-mechanism.

In contrast to the cool case B, case A at $\log \teff = 4.38$, behaves
differently: The only driving region (that is
$\diff \left(\oint_\Pi \diff\,\hat{q}\right) / \diff r > 0$)
encountered in this model star coincides with
$\diff\,\kappa_{T}/\diff\,r > 0$ inflicted by Rosseland-opacity's
$Z$-bump. The magnitude of thermal imbalance of the QHE model at the
evolutionary epoch of case A is much smaller than for case B so that
the relative contribution to $\oint_\Pi \diff\,\hat{q}$ of the star's
pulsation is larger.

By and large, the failed spatial matching of heat deposition in the
stellar envelope and the opacity peaks
(cf. Fig.~\ref{fig:M350Z02_CycleIntegratedHeat_and_StructuralProperties})
questions the role of the classical $\kappa$-effect to drive the
energetic pulsations that were encountered in the \MESA\,computations.
Additionally, given the rapid growth rates of the oscillatory
instabilities, it appears reasonable to search for signs of strange
modes. Strange modes, though, are not a monolithic phenomenon.
Depending on what qualifies as a strange mode, different processes can
lead to rapidly growing cyclic instabilities and to unusual period
evolution (compared with adiabatic pressure modes) as stellar
parameters vary \citep[e.g.][]{Glatzel1994,SBG98,Sonoi2014}. All
authors agree that strange-mode phenomena develop when the
radiation-pressure contribution to a star's structure is significant;
this usually goes along with stars sporting $L_\ast/M_\ast$ ratios
around and exceeding $10^{4}$. All the model stars of the model
sequences discussed in this report meet this criterion.
 
In the context of massive stars on or close to the main sequence, so
called "radiation-pressure dominated" (rpd) strange modes were found
to be relevant \citep[cf.][]{SBG98,Sonoi2014}.  Therefore, a
dissection of the momentum equation alone might already provide
insights:
\begin{equation}
	\ddot{r}  = -\frac{G m}{r^2} - 4\pi r^2 
	             \frac{\diff \left( \Pg + \Prad \right)}{\diff m} \,.
	\label{eq:momentum}
\end{equation}
Use the following shorthands for acceleration contributions on the right-hand side: 
$$ 
g_{\mathrm{grav}} = \frac{G m}{r^2}\, , \, \, 
g_{\mathrm{gas} } = - \frac{1}{\rho}\frac{\diff \Pg}{\diff r}
  \quad \mathrm{and} \quad 
g_{\mathrm{rad} } = - \frac{1}{\rho}\frac{\diff \Prad}{\diff r}  \,.
$$
Equation~\ref{eq:momentum} then writes compactly as
\begin{equation}
	\ddot{r} = -g_{\mathrm{grav}} + g_{\mathrm{gas}} + g_{\mathrm{rad}} \,. 
\end{equation}

\begin{figure}[h]
	\centering
	\includegraphics[width=0.49\textwidth]{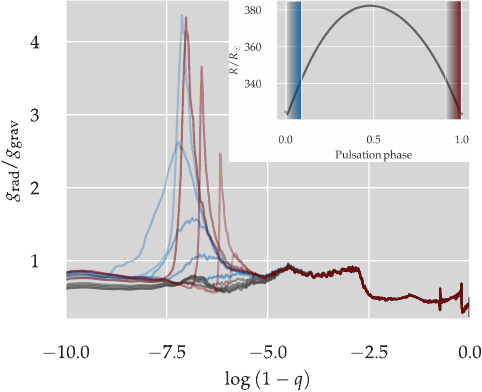}
	\hfill
	\includegraphics[width=0.49\textwidth]{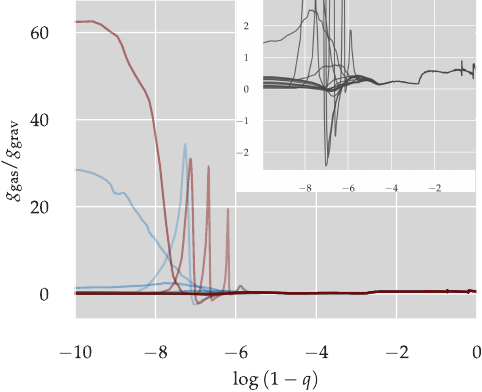}
	\caption{Superposition of spatial profiles of the ratios of  
			 $g_{\mathrm{rad}}$ (left) and  $g_{\mathrm{gas}}$ (right)
		  to $g_{\mathrm{grav}}$ at selected phases during 
		     a pulsation cycle of case-B model.
		    }
	\label{fig:M350Z02_LCLCycle3_phases_acclerations}
\end{figure} 
To learn how the different acceleration terms behave through a
pulsation cycle, Fig.~\ref{fig:M350Z02_LCLCycle3_phases_acclerations}
plots the profiles of the ratios of gas- and radiation-pressure
gradients relative to $g_{\mathrm{grav}}$ for selected phases during a
late pulsation cycle of case-B model.  The individual profiles had to
be numerically smoothed because apparently the tabular microphysics
used in \MESA\,inflicts spikes in the numerical evaluations of
$\diff \Prad / \diff r$ and $\diff \Pg / \diff r$, respectively.  The
profiles of the phases highlighted in blue and red color are those
clustered around successive minimum radii. The situation is
graphically highlighted in the inset of the left panel.  Except for a
small fraction of a pulsation period, $g_{\mathrm{rad}}$ is slightly
smaller than $g_{\mathrm{grav}}$ over the most of the star's envelope.
During maximum compression though, $g_{\mathrm{rad}}$ exceeds the
gravitational acceleration by up to a factor of four. The temporal
extension of this kick is color-coded on top of the radius curve
inserted on the upper right of the panel. This $g_{\mathrm{rad}}$
behavior is seen to be even aggravated by $g_{\mathrm{gas}}$, which is
displayed in right panel of
Fig.~\ref{fig:M350Z02_LCLCycle3_phases_acclerations}. Around minimum
radius again, $g_{\mathrm{gas}}$ can exceed gravitational acceleration
more than 60 times; this means that the star experiences a very
strong outward acceleration as it passes through maximum compression.
In a very narrow region, even in the mostly radiation dominated
envelopes of massive blue stars, the gas pressure and its gradient
can, under suitable circumstances, overwhelm that of
radiation-pressure. Evidently, it is in the mass-depth range $-7
< \log (1-q) < -9$ and during maximum compression that
$g_{\mathrm{gas}}$ dominates the resulting acceleration.  The inset in
the right panel of
Fig.~\ref{fig:M350Z02_LCLCycle3_phases_acclerations} zooms in
vertically to make visually accessible the behavior of
$g_{\mathrm{gas}}$ throughout the whole pulsation cycle: For most
phases of a cycle, $g_{\mathrm{gas}} / g_{\mathrm{grav}}$ remains
significantly smaller than unity in the pulsation-relevant envelope;
in very narrow layer (around $\log (1 - q) = -7$) the ratio can even
go negative, which would mean dynamical instability in absence of
radiation pressure and its gradient.

The acceleration spikes seen around minimum radius
(Fig.~\ref{fig:M350Z02_LCLCycle3_phases_acclerations}) are associated
with corresponding steep density gradients that build up during
maximum compression.  The physical reason in case B is the very sharp
opacity peak due to partial H/HeI ionization that develops temporally
close to the photosphere .  For hotter models, such as case A~--~which
also sports strong acceleration spikes~--~it is the narrow
sub-photospheric HeII partial ionization zone that causes them.

The acceleration phenomenon described above does not overlap spatially with 
the $\hat{q}$ gain regions. Therefore, stellar opacity plays a different 
role in the massive blue stars than in the canonical Eddington-type valve 
mechanism that acts in classical pulsators. In a Cephe\"{i}d model
with $6.1\,\msol$ (at $L/\lsol = 3377, \teff=5300$~K), 
as representative of classical pulsators, the acceleration imbalance of
$g_{\mathrm{gas}} / g_{\mathrm{grav}}$ was found to not exceed $1.5$ 
and the spatial maximum is confined to the main driving region of its pulsation;
the effect of $g_{\mathrm{rad}}$ remained negligible at all phases.  

\begin{figure}
	\centering
	\includegraphics[width=0.47\textwidth]{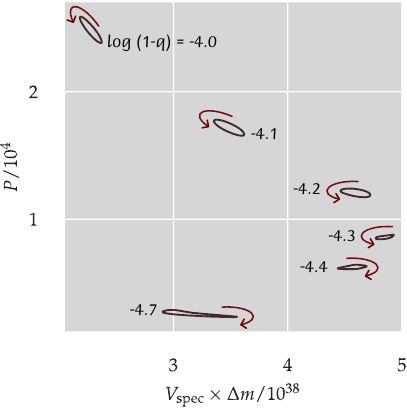}
	\hfill
	\includegraphics[width=0.47\textwidth]{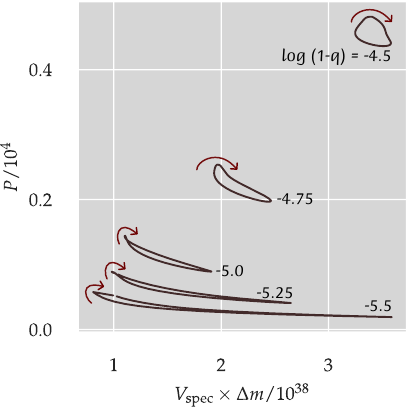}
	\caption{Well defined loops traced on the $P - V$ plane (in cgs units) of 
		selected mass layers over one late pulsation cycle of case-B model.
	}
	\label{fig:M350Z02_LCLCycle3_P-Rho-Loops}
\end{figure} 

\citet{Sonoi2014} pointed out that strange modes
show a characteristic spatial $\delta P - \delta \rho$ phase-shift in linear
theory. What do the nonlinear \MESA\,computations have to say in this respect? 
Along the lines of e.g. Fig.~9 in \citet{Christy1964}, 
$P(m_i) - V_{\mathrm{spec}}(m_i) \times \Delta m_i$ loops over a pulsation
cycle of selected mass shells are shown in 
Fig.~\ref{fig:M350Z02_LCLCycle3_P-Rho-Loops}. The quantity 
$V_{\mathrm{spec}}$ is the specific volume, 
i.e.  $V_{\mathrm{spec}} \equiv 1/ \rho$. The left panel of
Fig.~\ref{fig:M350Z02_LCLCycle3_P-Rho-Loops} shows
damping regions (loops are run through anti-clockwise) deeper
in the envelopes and then a transition to an excitation domain where
the loops are traversed clockwise ($\log (1 - q) < -4.4$). 
The right panel zooms in to the even more superficial layers
where pulsational driving persists.  Going from $\log (1 - q) < -4.0$
to $-4.4$, the orientation of the loops hint at a phase-shift of
pressure and density variation that changes such that it can be
interpreted as a $180$~degree shift if the temporal variation is
described with harmonic functions.
In contrast to the \citet{Sonoi2014} modeling, the loops derived from 
\MESA\,computations reorient themselves in the most superficial
regions of the star to again achieve maximum pressure around maximum density 
(cf. lower-left loops in the right panel of
Fig.~\ref{fig:M350Z02_LCLCycle3_P-Rho-Loops}).

\newpage
\centerline{\afgsection{5. WRAPPING IT UP FOR NOW}}

\MESA\,in hydrodynamical mode has already a reputation of being able to detect 
strong pulsational instabilities and with appropriate numerical adjustments 
to follow them over some time . This report presents first results that prove 
\MESA's successful picking up pulsations of massive blue stars. Their cyclic 
small-amplitude variability is observed e.g. in LBVs and theoretical studies
of linear and nonlinear pulsation analyses frequently attribute it to strange modes. 

In the instability domains found for $35, 45$, and $60 \msol$ model
stars in this report, most epochs did not reveal finite amplitude
pulsations.  Close to the instability boundaries though, cases were
found where a decent limit-cycle behavior developed that facilitated
cycle-integrated analyses to learn more about mode character.

The strangeness encountered in the pulsations of massive
blue stars was not only evident in fast amplitude growth, with 
e-folding times of the order of a few pulsation periods.
Also the excitation region of the pulsations is not coupled to the
opacity bumps in the same way as in classical pulsators.  
A satisfactory assignment to one of the families of strange modes 
\citep[cf.][]{SBG98} has not yet been achieved.
The overuse of concepts from linear theory in the 
interpretation of nonlinear processes poses a constant danger.
It seems clear though that the $\varkappa$-mechanism's way of 
action should be considered more broadly: 
For strange modes, the direct effect of the pressure gradient in the 
momentum equation appears to be decisive. For the situation of a 
radiation dominated stellar envelope,  Hideyuki Saio derived a 
limiting-case model~--~his rpd strange-mode class~--~which is dominated by the 
action of the \textit{radiation-pressure }gradient in the momentum 
equation \citep{SBG98}. The modulation of $g_{\mathrm{rad}}$ and its
partial temporal and spatial dominance over $g_{\mathrm{grav}}$ in massive 
blue stars pointed in direction of rpd modes. However, an even  much stronger
effect was exerted by very steep~--~opacity induced~--~gas-pressure 
gradients in the most superficial stellar layers. 
 
In classical pulsators on the other hand, \emph{energy}~absorption and
emission by the working substance~--~the gas~--~is essential, in close
analogy with the Carnot cycle of heat engines. The classical
$\varkappa$-mechanism as it operates say for Cephe\"ids and RR Lyrae
stars was therefore suggested to be renamed to $\varkappa$\textit{-energy
mechanism }(Lucy 1995, unpublished, see Appendix B).  For
strange-mode~--~like pulsations on the other hand, the effect of
strong and spatially narrowly confined opacity bumps that induce steep
radiation-pressure gradients might be referred to as
$\varkappa$\textit{-momentum mechanism }(Lucy 1995, unpublished).  The
current computations hint at an even broader domain of applicability
of the $\varkappa$-momentum mechanism as $g_{\mathrm{gas}}$
can \emph{locally }dwarf $g_{\mathrm{rad}}$ in otherwise radiation
dominated envelopes of blue massive stars.  Because it is the gradient
of a pressure gradient that is decisive for the $\varkappa$-momentum
mechanism, a dominance by magnitude of a pressure component is not
key. Nonetheless, the suggestion of a possibly vertical red edge of the instability 
region, as indicated by the $35$ and $45\,\msol$ evolutionary tracks (Fig.~\ref{fig:MassiveHRD}), points to the $\varkappa$-energy mechanism 
not being entirely negligible.

The regions of the stellar envelopes of massive blue stars where  
$g_{\mathrm{gas}} / g_{\mathrm{grav}}$ peaks are so superficial that the
involved thermal timescale is much shorter than the pulsation period. 
Hence, the classical $\varkappa$-mechanism (i.e. the $\varkappa$-energy mechanism) 
cannot kick in. Intuitively, this makes it at least plausible why 
the phase shifts between thermal and dynamical observables 
(Fig.~\ref{fig:M350Z02_VariousPhasePlots}) shrink or even disappear.

For high-energy pulsations as encountered in the blue massive stars
and even more so for the cool high-luminosity stars
\citep[cf.][]{Gautschy2023a,Gautschy2023b}, quantitative studies will need 
carefully modeled transitions to the circumstellar neighborhood to correctly 
account for effective pulsation-energy losses from the stellar surface. 
Furthermore, computational precautions must be implemented to include 
possible feedback to the stellar photosphere. 

One-dimensional dynamical stellar-evolution computations can only
paint an overly simplified picture of the behavior of a pulsating star
in complex circumstellar environments
\citep[cf.][]{Jiang2018, Goldberg2025, Bronner2025}.
Nevertheless, even with some persistent numerical deficiencies, 
useful insights into mode physics could be gained from the results achieved so far.
From this partial success, it seems worthwhile to invest further effort in  
\MESA\,to improve it beyond its already unparalleled capabilities.
To compute broader classes of nonlinear radial pulsators with a
dynamical stellar-evolution code, the numerical requirements to fulfill physical 
conservation laws are even higher than what can be achieved currently.

\bigskip

\textsc{Acknowledgments} NASA's Astrophysics Data System continued to serve
as a most reliable, indispensable virtual library. 
The model-star sequences discussed in this report could not have been 
computed without the \MESA\,~soft\-ware-instrument~--~used here in version
r$21.12.1$~--~\citep{Paxton2019}. 
Post-processing \MESA{\tt star}\,models and plotting benefited from
Warrick Ball's \href{https://github.com/warrickball/tomso}{\texttt{tomso}}
module, \texttt{numpy} \citep{harris2020array}, \texttt{astropy}
\citep{astropy2022}, \texttt{scipy} \citep{2020SciPy}, 
and \texttt{matplotlib} \citep{Hunter2007}, respectively.
Hideyuki Saio's judicious comments are much appreciated.

\newpage

\centerline{\afgsection{APPENDIX A}}

The computational setup is based on \MESA\,version r$21.12.1$. The
model stars of the sequences referred to in this report were evolved
conservatively in dynamical mode from the ZAMS: The acceleration term
in the momentum equation was switched on by setting
\texttt{change\_v\_flag = .true.} and \texttt{new\_v\_flag = .true.} in the 
\texttt{inlist} file.
Regarding the choice of the energy-equation form and the quality of 
its conservation:
\texttt{energy\_eqn\_option = 'dedt'} and 
\texttt{use\_gold\_tolerances = .true.} were requested. 

The Schwarzschild criterion regulated convective stability, 
the mixing length was set to $1.8$ pressure 
scale-heights. Semiconvection and thermohaline mixing were ignored. 
Convective overshooting was assumed to operate in exponential mode with 
\texttt{overshoot\_f = 0.012, overshoot\_f0 = 0.002}.  It
is worthwhile to notice that the \texttt{inlist} entries that specify
the detailed microphysics of the model stars need not be finely tuned
for pulsations to develop. Any choices as found e.g. in \MESA 's
massive-stars \texttt{test\_suites} are likely fine as initial trials. 
All that counts is if the \texttt{inlist} choices 
take a model star into the relevant domain on the HR plane
(cf. Fig.~\ref{fig:MassiveHRD}) while $L_\ast/M_\ast \gtrsim 10^4$.

The temporally high-resolution runs restarted from \texttt{photos}
models imposing a maximum timestep,
\texttt{max\_years\_for\_timestep}, between $10^{-3} - 10^{-4}$. 
In contrast to the earlier dynamical models with very short 
timesteps (post~-~AGB stars and R~CrB stars), no regridding was 
necessary or it was even counterproductive to induce numerical noise 
that was intended to help to start pulsations.
The choices of \texttt{max\_years\_for\_timestep} finally adopted in model sequences 
with saturating pulsations were usually dictated by the aim to arrive
at limit-cycle like behavior. Hence, the steady pulsations must be considered
substantially affected by the remaining numerical dissipation in \MESA.
This approach worked only close to the instability edges. Deep within the
instability region, pulsation growth was so strong that no
\texttt{max\_years\_for\_timestep} limiter could be found whatsoever to 
simultaneously saturate \emph{and }temporally resolve the pulsation. 

\newpage

\centerline{\afgsection{APPENDIX B}}

This work concludes with a brief personal recollection of the
motivation behind it: As far as I know, it was Leon Lucy
who first decomposed the $\varkappa$-mechanism into a gas and a radiation
contribution  in the mid 1990s when he tried to understand LBVs.  
I heard him lecturing on it in the fall of 1995. 
Unfortunately, Leon never wrote down his research for publication.
\emph{I }forgot about it, mainly because I failed to fully understand 
Leon~--~as so often. In 1998, for the \citet{SBG98} paper, Hideyuki
Saio derived linear pulsation models for limiting cases that yielded strange modes. 
The pulsational instability of his rpd class feeds essentially on the
$\varkappa$-momentum mechanism. At that time I failed to connect the dots 
and I did not realize that Hideyuki's independent derivation matched Leon
Lucy's vision. Some years later, around 2010, while clearing out a closet, I 
stumbled upon notes on the 1995 talk. Ernst Dorfi and I then planned to 
take a closer look at Leon's hydrodynamics; we set out to redo 
(with the latter's blessings) some of the nonlinear computations 
with the~--~as we presumed~--~more versatile Viennese RHD code. 
But we got nowhere, and the project fizzled out, like so many others 
of its kind at that time. Serendipitously, I was granted
yet another chance~--~with \MESA\,~--~and I decided to make one
final attempt. This note reports on how far I got. The rest be left to others.

\bigskip\bigskip

\bigskip\bigskip

\centerline{\large{\textsc{To the memory of}}}

\medskip

\centerline{\large{\textsc{Leon Lucy, Ernst Dorfi, and Bill Paxton}}}

\bibliographystyle{aa}
\bibliography{/home/alfred/StarDude/Librarium/StarBase}

\end{document}